\documentclass[%
pre,reprint,longbibliography,onecolumn,
 amsmath,amssymb,
 aps,
]{revtex4-2}

\usepackage{bm}
\usepackage{graphicx}
\usepackage{enumitem}   
\usepackage{cleveref}
\graphicspath{{graphics/}}
\usepackage{float}
\usepackage{xcolor}

\begin{document}

\widetext

\title{Random walks with long-range memory on networks}


\author{Ana Gabriela Guerrero-Estrada$^{1}$}
\email{anaguerrero@estudiantes.fisica.unam.mx}
\author{Alejandro P. Riascos$^{2}$}
\email{alperezri@unal.edu.co}
\author{Denis Boyer$^{1}$}%
 \email{boyer@fisica.unam.mx}
\affiliation{%
 $^1$Instituto de F\'isica, Universidad Nacional Aut\'onoma de M\'exico, Mexico City 04510, Mexico
}%
\affiliation{%
$^2$Departamento de F\'\i sica, Universidad Nacional de Colombia,
Bogot\'a, Colombia}

\date{\today}

\begin{abstract}
We study an exactly solvable random walk model with long-range memory on arbitrary networks. The walker performs unbiased random steps to nearest-neighbor nodes and intermittently resets to previously visited nodes in a preferential way, such that the most visited nodes have proportionally a higher probability to be chosen for revisit. The occupation probability can be expressed as a sum over the eigenmodes of the standard random walk matrix of the network, where the amplitudes slowly decay as power-laws at large time, instead of exponentially. The stationary state is the same as in the absence of memory and detailed balance is fulfilled. However, the relaxation of the transient part becomes critically self-organized at late times, as it is dominated by a single power-law whose exponent depends on the second largest eigenvalue and on the resetting probability. 
We apply our findings to finite networks such as rings, complete graphs, Watts-Strogatz and Barab\'asi-Albert networks, and to Barbell and comb-like graphs. Our study  
could be of interest for modeling complex transport phenomena, such as human mobility, epidemic spreading, or animal foraging.
\end{abstract}

\maketitle

{\bf Random walks on networks are rather well-known processes and mostly studied in the realm of the Markov hypothesis. The properties of non-Markov random walks on networks are much less understood, especially if their memory is long-ranged. Here we exactly solve a random walk model with long-range memory on arbitrary networks. The model consists of a standard nearest-neighbor random walk with stochastic relocation jumps to nodes visited in the past. This relocation process is preferential, i.e., the likelihood to revisit a certain node depends linearly on the total amount of time spent by the walker on this node, an assumption consistent with empirical observations on human and animal mobility. We show that this model exhibits a slow, anomalous relaxation toward the steady state at late times, in the form of an inverse power-law whose exponent is determined analytically.}

\section{Introduction}
Dynamical processes on networks occur in various domains of sciences, ranging from microscopic transport processes in cells or the neuronal activity of the brain, to the modeling of traffic in cities and the evolution of  epidemics \cite{VespiBook,barabasi2016book,newman2018}. In many contexts, an important question is how to quantify the capacity of transport in a given network. Indicators that serve to characterize spreading dynamics can be obtained by studying random walk processes on the underlying network, for instance, the evolution of the occupation probability, the stationary distributions or the mean first-passage times, which measure the average times needed by a walker to reach a specific target site for the first time \cite{Kemeny,Montroll1956Random,Hughes,masuda2017random,Lovasz1996}. Over the past decades, researchers have extensively investigated random walks on a variety of network structures and studied the impact of different strategies where the walker may use local and non-local information about the network \cite{noh2004random,masuda2017random,ReviewJCN_2021}. The results have demonstrated that both the structure of the network and the random walk strategy significantly affect the spreading dynamics.
\\[2mm]
The mounting understanding of random walks on networks has led to results of wide applicability; for example, the development of algorithms in data science, such as PageRank for Internet searches \cite{brin1998,BlanchardBook2011}. Despite its success in various fields, random walk modeling generally assumes that the processes are Markovian, i.e., their evolution from a given node does not depend on the past. This framework does not apply to processes where memory effects are important, for instance when the walker makes movement decisions based on the nodes that it has already visited.
Unfortunately, our current understanding of non-Markovian random walks on networks is extremely limited in comparison to that of their memory-less counterparts. The difficulties are mainly due to the lack of an appropriate theoretical framework to handle memory terms in master equations. Random walks with one-step memory (i.e., close to Markov) have been studied on complex networks, showing that they can detect other types of communities than simple random walks \cite{rosvall2014memory}, be more accurate to evaluate diffusive flows through modules \cite{lambiotte2015effect} or be locally efficient for navigation  \cite{basnarkov2020random,cao2021one}. Random walks with longer, $n-$step memory can be designed to complete a search task on regular lattices more rapidly than the standard random walk \cite{Rieger_PRL_2021}. Simulations have also shown that self-avoiding random walkers explore networks efficiently \cite{kim2016network}, in particular when their memory is reset to zero from time to time \cite{Colombani_2023}. Self-avoiding processes on networks have motivated mean-field models \cite{lopez2012model} and new methods for quantifying network robustness \cite{valente2022non} or detecting communities \cite{de2018community}.
\\[2mm]
Recent analytical and numerical studies have considered a class of models where, in contrast, the walk tends to revisit previous locations. An example is an ordinary random walk that resets with a given probability directly to a fixed node (or to one of several fixed nodes) from anywhere in the network \cite{ResetNetworks_PRE2020,Reset_MultipleNodes_2021,wang2021random,ye2022random,zelenkovski2023random}. This is analogous to remembering specific locations that are revisited from time to time and from which the process starts anew. Resetting processes to a single site (typically, the starting position) have been extensively studied over the past decade in one-dimensional systems in a variety of contexts, such as diffusion, the kinetics of biochemical reactions or animal foraging \cite{evans2011diffusion,
evans2011bdiffusion,evans2020stochastic,gupta2022stochastic,pal2017first,belan2018restart,evans2014diffusion,boyer2014random,giuggioli2019comparison,pal2020search,reuveni2014role,rotbart2015michaelis,pal2019landau,campos2015phase,
besga2020optimal,tal2020experimental,blumer2024combining}. Resetting often expedites the search time in random search problems. These processes also violate detailed balance and evolve at late times toward a non-equilibrium steady state. Recently, Bae {\it et al.} introduced a non-Markov exploration on networks with long range memory based on a truly non-Markovian resetting protocol, where a walker can either perform a standard random walk step or reset to a previously visited node, such that all the visited nodes have the same probability to be chosen \cite{bae2022unexpected}. This strategy turns out to accelerate target searches taking place on lollipop-like networks, although it is probably advantageous in other situations, too. 
\\[2mm]
Inspired by resetting processes, a class of exactly solvable models with long range memory have been studied over the past decade in one spatial dimension. A basic model consists of a walker on a regular lattice that either performs a nearest-neighbor (n.n.) random step or, with some constant probability, relocates (or resets) to a site visited in the past \cite{boyer2014random}. Resetting is preferential in this model, i.e., not all the visited nodes have the same probability to be chosen during a relocation event: instead, a particular site is chosen with a probability proportional to the total amount of time spent there by the walker since the beginning of the walk. Hence, more familiar places are more likely to be revisited, a behavior which is commonly observed in human \cite{RiascosMateosPlos2017,song2010modelling,pappalardo2016human,wang2019extended,schlapfer2021universal,cabanas2023human} or animal mobility \cite{boyer2014random,merkle2014memory,vilk2022phase,Ranc_PNAS_2021}. On infinite lattices, due to the reinforcement of the most visited sites, this model generates ultra-slow, logarithmic diffusion with an occupation probability that tends to a Gaussian. Extensions to continuous space or continuous time situations have been considered, where the above resetting protocol with preferential memory is applied to other underlying Markov processes such as Brownian motion \cite{boyer2017long}, L\'evy flights \cite{boyer2016slow}, continuous time random walks \cite{campos2019recurrence, maso2019anomalous}, active particles \cite{boyer2024active}, or to cases with memory decay \cite{boyer2014solvable,boyer2017long}. Considering a wide class of memory kernels and resetting time distributions, Refs. \cite{mailler2019random,boci2024central} have rigorously proven central and local limit theorems for these processes when the underlying process is in ${\mathbb R}^d$. A large deviation principle has also been established \cite{boci2023large,boyer2024active}. 
\\[2mm]
When a preferential relocation process of the type described above is restricted to a finite number of sites \cite{campos2019recurrence} or confined in space by an external potential \cite{boyer2024powerlaw}, the process admits a steady state distribution. For a confined Brownian particle subject to preferential relocation, the relaxation toward the steady state was found to be anomalous \cite{boyer2024powerlaw}. It is well-known that standard Brownian particles in confining potentials relax exponentially at late times toward the Gibbs-Boltzmann distribution. If preferential memory is switched on, in contrast, relaxation toward the stationary state lacks a characteristic time-scale and follows an inverse power-law. The exponent of this power-law depends on the resetting rate and on the lowest relaxation rate of the memory-less diffusion process in the same potential \cite{boyer2024powerlaw}.
\\[2mm]
In this paper, we extend the basic lattice random walk model with long-range memory of \cite{boyer2014random} to arbitrary finite networks, where a steady state distribution exists in principle. This model can be studied analytically because an exact master equation for the occupation probability can be written under the non-Markovian dynamics. This master equation combines the hopping probabilities to n.n. nodes and a term including the effects of memory. The solutions can be found by using the eigenvalues and eigenvectors of the transition matrix of the standard random walk, although the time evolutions differ markedly. 
We find that the occupation probability slowly decays toward its steady state as a power-law at large times, instead of the usual exponential decay that characterizes Markovian dynamics. 
%
The paper is organized as follows. In section \ref{sec:model}, we define the model and summarize the main analytical results. We provide a derivation of these results in sections \ref{sec:disc}
and \ref{sec:cont}, which are dedicated to the discrete and continuous time formulations, respectively. In section \ref{sec:examples}, we apply the results to several representative network topologies, such as complete graphs, rings, Watts-Strogatz and Barab\'asi-Albert complex networks, Barbell graphs and combs. We conclude in section \ref{sec:concl}.

\section{Model definition and summary of main results}\label{sec:model}

\begin{figure}[t]
\centering
\includegraphics[width=0.35\textwidth]{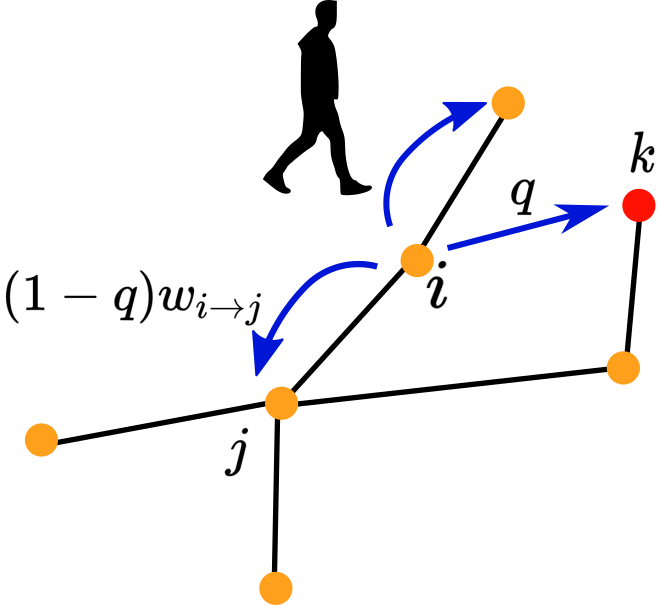}
\caption{Model analyzed in the present study. A random walker on a finite, single component network hops randomly to a nearest-neighbor node with probability $1-q$ or reset to a previously visited node with the complementary probability $q$. In the latter case, the probability to reset to a particular node $k$ (among the visited nodes) is proportional to the time spent on $k$ since the beginning of the process. The transition probabilities in the random walk mode are defined by the matrix elements $w_{i \rightarrow j}\equiv w_{ij}$.}
\label{fig.model}
\end{figure}

We start by defining the rules of the model in discrete time and summarize the main analytical results, before discussing the continuous-time version. During a time-step $t\rightarrow t+1$, a walker located at the node $i$ of an arbitrary single-component network of $N$ nodes can perform one of the two following actions: $(i)$ with probability $1-q$, where $q$ is a parameter in the interval $[0,1]$, the walker hops to a n.n. node chosen randomly with equal probability; $(ii)$ with the complementary probability $q$, the walker resets directly to a previously visited node $j$. In rule $(ii)$, the node $j$ is chosen, among all the visited nodes, with probability proportional to the total number ($m_j$) of visits it has received since the initial time (see Fig. \ref{fig.model}). Therefore, the action $(ii)$ introduces long-range memory in the process and, through this rule, frequently visited nodes are more likely to be revisited than rarely occupied sites. The standard random walk is recovered by setting $q=0$.
\\[2mm]
The main quantity of interest here is $P_{ij}(t)$, the probability that the walker occupies the node $j$ after $t$ steps, having started from node $i$ at $t=0$. This occupation probability can be decomposed into a stationary part and a transient part
\begin{equation}
P_{ij}(t)=P_{j}^{({\rm st})}+P_{ij}^{({\rm tr})}(t),
\label{sum_decomp}
\end{equation}
where $P_{ij}^{({\rm tr})}(t)\rightarrow 0$ as $t\rightarrow\infty$. The stationary part satisfies detailed balance and is the same as for the memory-less walk \cite{noh2004random}, 
\begin{equation}
P_{j}^{({\rm st})}=\frac{k_j}{\sum_{n=1}^N k_n},
\label{sum_sta}
\end{equation}
where $k_j$ is the degree of node $j$. The transient part of the distribution is know exactly through its characteristic function $\widetilde{P}_{ij} (z)\equiv\sum_{t=0}^{\infty}z^t P_{ij}(t)$, given by the rather long expression in Eq. (\ref{charPij}). The large time decay of $P_{ij}^{({\rm tr})}(t)$ can be deduced from this result and the leading behavior takes the form of an inverse power-law,
\begin{equation}
P_{ij}^{({\rm tr})}(t)\propto \frac{1}{\Gamma(1-b_2(q))}\, t^{-b_2(q)},
\label{sum_tr_disc}
\end{equation}
where $\Gamma(\cdot)$ is the Gamma function. Apart from depending on $q$, the exponent $b_2(q)$ also involves the spectral gap of the network, i.e., the second largest eigenvalue $\lambda_2$ of the $N\times N$ transition probability matrix ${\mathbf W}$ of the random walk {\it without} memory on the same network (see Section \ref{sec:disc} below). Its expression is given by
\begin{equation}\label{sum_b2}
b_2(q)=\frac{(1-q)(1-\lambda_2)}{1-(1-q)\lambda_2}.
\end{equation}
This exponent is smaller than 1 and positive, since $\lambda_2<1$. Clearly, the anomalous relaxation in Eq. (\ref{sum_tr_disc}) contrasts with the exponential decay $P_{ij}^{({\rm tr})}(t)|_{q=0}\sim \lambda_2^t$ that characterizes the standard Markov random walk at late times on the network \cite{Lovasz1996,masuda2017random}. Here, the stationary state is approached very slowly, a property explained by the fact that, due to memory, the walker often revisits sites that are in the vicinity of the starting node. A non-trivial feature is the absence of a characteristic relaxation time, which renders the system \lq\lq critical" for all values of $q\in(0,1)$. In other words, the process always self-organizes into a critical state (with a non-universal exponent) as a consequence of self-attraction and of the discrete spectrum of the finite network.
\\[2mm]
In the continuous time version of this model, a walker standing at node $i$ can move to a n.n. node with rate $\gamma$ or reset to a node visited in the past with rate $r$. In the latter case, a given previous node is chosen with probability proportional to the total amount of time spent on that node by the walker since $t=0$, similarly to the preferential rule of the discrete-time model. The occupation probability in this continuous setting is also decomposed as in Eq. (\ref{sum_decomp}), with the same stationary part (\ref{sum_sta}). For the transient part, an explicit time-dependent expression can be obtained, valid for all $t$,
\begin{equation}
P_{ij}^{\rm (tr)}(t)=\sum_{\ell=2}^N M\left( \frac{\gamma(1-\lambda_{\ell})}{\gamma(1-\lambda_{\ell})+r},\, 1,\, -[\gamma(1-\lambda_{\ell})+r]t\right) \langle i|\phi_{\ell}\rangle\langle\bar{\phi}_{\ell}|j\rangle ,
\label{sum_tr_cont}
\end{equation}
where, in Dirac's notation, $\{|\phi_{\ell}\rangle,\,\langle\bar{\phi}_{\ell}|\}$ represent the right and left eigenvectors, respectively, of the matrix ${\mathbf W}$ of the random walk {\it without} memory, with eigenvalues $\{\lambda_{\ell}\}$, while $|i\rangle$ is the canonical vector of unit norm with all its components 0 except the $i$-th one
(see Section \ref{sec:disc}). The function $M(a,b,z)$ in Eq. (\ref{sum_tr_cont}) is the confluent hypergeometric function $_1F_1$, or Kummer's function, defined through its power series \cite{abramowitz1965handbook}
\begin{equation}
M(a,b,z)= 1 + \frac{a}{b}\, z+ \frac{a(a+1)}{b(b+1)}\, \frac{z^2}{2!}+ \frac{a(a+1)(a+2)}{b(b+1)(b+2)}\, \frac{z^3}{3!}+ \ldots\, .
\label{M_def.1}
\end{equation}
By ordering the eigenvalues from largest to lowest, or $\lambda_1=1\ge\lambda_2\ge\lambda_3\ge\ldots\ge\lambda_N$, the transient part given by Eq. (\ref{sum_tr_cont}) is dominated by the single mode $\ell=2$ at late times. Due to the algebraic decay of $M(a,b,-z)$ at large $z$ for $a<b$, the relaxation toward the steady state is found to be critical, like in the discrete time model, 
\begin{equation}
P_{ij}^{(\rm st)}(t)\propto \frac{1}{\Gamma(1-\theta_2(r))}\, t^{-\theta_2(q)},
\end{equation}
where
\begin{equation}\label{theta2}
\theta_2(r)=\frac{\gamma(1-\lambda_{2})}{\gamma(1-\lambda_{2})+r}\, .
\end{equation}
Again, this decay is qualitatively very different from the much faster exponential relaxation of the standard Markov continuous time random walk, which is recovered by setting $r=0$ in Eq. (\ref{sum_tr_cont}). In that case, the $M$ functions simplify to $e^{-\gamma(1-\lambda_{\ell})t}$ and $P_{ij}^{\rm (tr)}(t)|_{q=0}\sim  e^{-\gamma(1-\lambda_2)t}$ at large time. An important feature of both the discrete and continuous time formulations summarized above, is the fact that the occupation probability of the process with memory can be expressed in terms of the spectral properties of the random walk in the absence of memory.

\section{Discrete time formulation}\label{sec:disc}

We now derive these results, starting with the discrete-time dynamics. The evolution of the occupation probability $P_{ij}(t)$ defined previously can be described by the following master equation,
\begin{equation}\label{master1}
P_{ij}(t+1)=(1-q)\sum_{m=1}^{N}P_{im}(t)w_{m j}+\frac{q}{t+1}\sum_{t'=0}^t P_{ij}(t'),
\end{equation}
with the initial condition $P_{ij}(t=0)=\delta_{ij}$.
In the first term, 
$w_{ij}$ defines the transition matrix ${\mathbf W}$ and represents the probability of hopping from $i$ to $j$ in one time step for the standard n.n. Markov random walk on the network. This process is unbiased, thus one has $w_{ij}=A_{ij}/k_i$, where $A_{ij}=1$ if $i$ and $j$ are connected and $0$ otherwise, whereas $k_i=\sum_{l=1}^N A_{il}$ is the number of nodes to which $i$ is connected. The second term in Eq. (\ref{master1}) represents the probability of revisiting node $j$ by the use of memory: the sum over $t'$ is $\langle m_j\rangle$, the number of previous visits received by $j$ up to time $t$ averaged over all the possible trajectories starting from $i$. To obtain a probability, this sum is normalized by the total number of visits $t+1$, which includes the initial time $t=0$. An alternate interpretation of the linear preferential memory process is as follows: choose an integer time $t'$ randomly and uniformly in the interval $[0,t]$, i.e. with probability $1/(t+1)$, and reset the walker to the position it occupied at time $t'$. (See also ref. \cite{boyer2014solvable} for a step-by-step derivation of the memory term.) An important property of Eq. (\ref{master1}) is that it is closed: it only involves the single-time distribution function $P_{ij}(t)$ and not high-order, multiple-time functions.
\\[2mm]
Using Dirac's notation, we denote as $\{|i\rangle\}_{i=1,\ldots ,N}$ the canonical basis of $\mathbb{R}^N$, where $|i\rangle=(0,\ldots ,0,1,0,\ldots 0)^T$ has vanishing components except the $i$-th one, while $\langle i|$ is the transposed of $|i\rangle$. In the following, we look for solutions of Eq. (\ref{master1}) in the basis of the right or left eigenvectors of ${\mathbf W}$, denoted as $|\phi_\ell\rangle$ and $\langle \bar{\phi}_{\ell}|$, respectively. Namely, we have
\begin{align}
{\mathbf W}|\phi_{\ell}\rangle&=\lambda_{\ell}|\phi_{\ell}\rangle\\
\langle\bar{\phi}_{\ell}|{\mathbf W}&=\lambda_{\ell}\langle\bar{\phi}_{\ell}|,
\end{align}
with $\langle\bar{\phi}_{\ell'}|\phi_{\ell}\rangle=\delta_{\ell'\ell} $ 
and $\sum_{\ell=1}^N |\phi_{\ell}\rangle\langle\bar{\phi}_{\ell}|=\mathbb{I}$ where $\mathbb{I}$ denotes the $N\times N$ identity matrix. We recall that the eigenvalues $\lambda_{\ell}$ are ordered from largest to lowest, $\lambda_1=1>\lambda_2\geq\ldots\geq\lambda_N\ge-1$. The spectral form of the transition matrix reads
\begin{equation}\label{W_spect}
	\mathbf{W}=\sum_{\ell=1}^N\lambda_{\ell}\left|\phi_{\ell}\right\rangle\left\langle\bar{\phi}_{\ell}\right|,
\end{equation}
with elements $w_{ij}=\sum_{\ell=1}^N	\lambda_\ell \left\langle i|\phi_{\ell}\right\rangle\left\langle\bar{\phi}_{\ell}|j\right\rangle$. In a similar way, we assume the following representation for the occupation probabilities $P_{ij}(t)$
\begin{equation}\label{Pij_spect}
	P_{ij}(t)=\sum_{\ell=1}^N c_{\ell}(t;q)\left\langle i|\phi_{\ell}\right\rangle\left\langle\bar{\phi}_{\ell}|j\right\rangle.
\end{equation}
The coefficients $c_{\ell}(t;q)$ are real and capture the temporal evolution of  $P_{ij}(t)$, where we have made explicitly their dependence on the memory parameter $q$. Note that $|\phi_{\ell}\rangle$, $\langle\bar{\phi}_{\ell}|$ and $\lambda_\ell$ {\em do not} depend on $q$ as they are properties of the standard random walk on the network.
\\[2mm]
Substituting Eqs. (\ref{W_spect}) and (\ref{Pij_spect})  into Eq. (\ref{master1}), one obtains
\begin{eqnarray}\label{master2}
	\sum_{\ell=1}^N c_{\ell}(t+1;q)
\left\langle i|\phi_{\ell}\right\rangle\left\langle\bar{\phi}_{\ell}|j\right\rangle
	&=&	(1-q)\sum_{m=1}^N \sum_{\ell=1}^N c_{\ell}(t;q)
\left\langle i|\phi_{\ell}\right\rangle\left\langle\bar{\phi}_{\ell}|m\right\rangle
\sum_{s=1}^N	\lambda_s \left\langle m|\phi_{s}\right\rangle\left\langle\bar{\phi}_{s}|j\right\rangle\nonumber\\
&+&\frac{q}{t+1}\sum_{t'=0}^t \sum_{\ell=1}^N c_{\ell}(t';q)\left\langle i|\phi_{\ell}\right\rangle\left\langle\bar{\phi}_{\ell}|j\right\rangle.
\end{eqnarray}
However, for the first term in the right hand side of Eq. (\ref{master2}),
\begin{eqnarray}
\sum_{m=1}^N \sum_{\ell=1}^N c_{\ell}(t;q)
\left\langle i|\phi_{\ell}\right\rangle\left\langle\bar{\phi}_{\ell}|m\right\rangle
\sum_{s=1}^N	\lambda_s \left\langle m|\phi_{s}\right\rangle\left\langle\bar{\phi}_{s}|j\right\rangle&=&
\sum_{\ell=1}^N c_{\ell}(t;q)\left\langle i|\phi_{\ell}\right\rangle\sum_{s=1}^N
\lambda_s \left\langle\bar{\phi}_{s}|j\right\rangle\sum_{m=1}^N\left\langle\bar{\phi}_{\ell}|m\right\rangle\left\langle m|\phi_{s}\right\rangle\nonumber\\
&=&\sum_{\ell=1}^N c_{\ell}(t;q)\left\langle i|\phi_{\ell}\right\rangle\sum_{s=1}^N
\lambda_s \left\langle\bar{\phi}_{s}|j\right\rangle\delta_{\ell s}\nonumber\\
&=&\sum_{\ell=1}^N c_{\ell}(t;q)\lambda_{\ell}\left\langle i|\phi_{\ell}\right\rangle\left\langle\bar{\phi}_{\ell}|j\right\rangle.
\end{eqnarray}
Therefore, Eq. (\ref{master2}) takes the form,
\begin{equation}
\sum_{\ell=1}^N \left\langle i|\phi_{\ell}\right\rangle\left\langle\bar{\phi}_{\ell}|j\right\rangle
\left[c_{\ell}(t+1;q)-(1-q)\lambda_\ell c_{\ell}(t;q)-\frac{q}{t+1}\sum_{t'=0}^t  c_{\ell}(t';q)\right]=0.
\end{equation}
As the equality holds for all $i$ and $j$, we obtain an equation for the coefficient $c_{\ell}(t;q)$ of mode $\ell$,
\begin{equation}\label{cm_master}
	c_{\ell}(t+1;q)=(1-q)\lambda_{\ell} c_{\ell}(t;q)+\frac{q}{t+1}\sum_{t^\prime=0}^t c_{\ell}(t^\prime;q).
\end{equation}
The recursive equation (\ref{cm_master}) has to be solved with the initial condition
\begin{equation}\label{ini_c_ell}
c_{\ell}(t=0;q)=1,
\end{equation}
which ensures that $P_{ij}(t=0)=\delta_{ij}$.
\\[2mm]
In the particular case $q=0$, Eq. (\ref{cm_master}) is simply solved as $c_{\ell}(t;q=0)=\lambda_{\ell}^t$, recovering the well-known standard random walk result \cite{masuda2017random,ReviewJCN_2021}. Since $\lambda_{\ell=1}=1$, the mode $\ell=1$ corresponds to the stationary part of the occupation probability, while the modes with $\ell\ge2$ are transient. Hence,
\begin{equation}\label{Pij_markovian}
	\left.P_{ij}(t)\right|_{q=0}=P_j^{\rm (st)}+\sum_{\ell=2}^N
	\lambda_{\ell}^t \langle i|\phi_{\ell}\rangle \langle\bar{\phi}_{\ell}|j\rangle,
\end{equation}
where
\begin{equation}
P_j^{\rm (st)}=\langle i|\phi_{1}\rangle \langle\bar{\phi}_{1}|j\rangle
\label{sta2}
\end{equation}
is equivalent to Eq. (\ref{sum_sta}).
In the long time limit, the transient part is dominated by the second largest eigenvalue,
\begin{equation}
\left.P_{ij}^{\rm (tr)}(t)\right|_{q=0}\simeq \lambda_2^t\,
\langle i|\phi_{2}\rangle \langle\bar{\phi}_{2}|j\rangle.
\end{equation}
To analyze the case $q>0$, let us introduce the generating function, defined as $\tilde{f}(z)\equiv \sum_{t=0}^\infty z^t f(t)$, and let us apply it to Eq. (\ref{cm_master}). One obtains
\begin{equation}
\frac{1}{z}\left[\tilde{c}_{\ell}(z;q)-c_{\ell}(t=0;q)\right]=(1-q)\lambda_{\ell}\tilde{c}_{\ell}(z;q)+\frac{q}{z}\int_0^z\frac{\tilde{c}_{\ell}(u;q)}{1-u}du.
\end{equation}
where we have used the identity $\frac{z^t}{t+1}=\frac{1}{z}\int_0^z u^tdu$ and performed the sums over $t$ and $t'$. Since $c_{\ell}(t=0;q)=1$, 
\begin{equation}\label{cm_master_z}
	\tilde{c}_{\ell}(z;q)=1+(1-q)\lambda_{\ell}z \tilde{c}_{\ell}(z;q)+q\int_0^z\frac{\tilde{c}_{\ell}(u;q)}{1-u}du.
\end{equation}
Taking the derivative of Eq. 
(\ref{cm_master_z}) with respect to $z$ gives
\begin{equation}
\frac{d \tilde{c}_{\ell}(z;q)}{dz}=(1-q)\lambda_{\ell} \tilde{c}_{\ell}(z;q)+(1-q)\lambda_{\ell} z \frac{d \tilde{c}_{\ell}(z;q)}{dz}+ q\frac{\tilde{c}_{\ell}(z;q)}{1-z}. 
\end{equation}
or
\begin{equation}
 \frac{d \tilde{c}_{\ell}(z;q)}{dz}=\left[  \frac{q+(1-q)\lambda_{\ell}(1-z)}{1-(1-q)\lambda_{\ell} z} \right]\frac{\tilde{c}_{\ell}(z;q)}{1-z}. 
\end{equation}
This first-order differential equation is solved as
\begin{equation}
\tilde{c}_{\ell}(z;q)=\exp\left\{-[1-b_{\ell}(q)]\ln(1-z)-b_{\ell}(q)\ln[1-(1-q)\lambda_{\ell} z]\right\},
\end{equation}
where we have again used the initial condition, which imposes $\tilde{c}_{\ell}(z=0;q)=1$. In the above expression, $b_{\ell}(q)$ is given by
\begin{equation}\label{ab} 
b_{\ell}(q)=\frac{(1-q)(1-\lambda_{\ell})}{1-(1-q)\lambda_{\ell}}.
\end{equation}
This solution can be recast in a more compact way as,
\begin{equation}\label{c_m_tilde}
\tilde{c}_{\ell}(z;q)=(1-z)^{-(1-b_{\ell}(q))}[1-(1-q)\lambda_{\ell} z]^{-b_{\ell}(q)}.
\end{equation}
The exact expression for the generating function of the occupation probability is deduced from Eqs. (\ref{Pij_spect}) and (\ref{c_m_tilde}),
\begin{equation}\label{charPij}
\widetilde{P}_{ij}(z)=\frac{\langle i|\phi_{1}\rangle \langle\bar{\phi}_{1}|j\rangle}{1-z}+\sum_{\ell=2}^N 
(1-z)^{-(1-b_{\ell}(q))}[1-(1-q)\lambda_{\ell} z]^{-b_{\ell}(q)}
\left\langle i|\phi_{\ell}\right\rangle\left\langle\bar{\phi}_{\ell}|j\right\rangle,
\end{equation}
where we have used the fact that $b_{\ell=1}(q)=0$ for all $q$, from Eq. (\ref{ab}).
We recognise in the first term the generating function of a constant, hence the stationary state $P_j^{\rm(st)}$ is still given by Eq. (\ref{sta2}) or (\ref{sum_sta}), like in the memory-less process.
\\[2mm]
For the other modes $\ell\ge2$, Eq. (\ref{ab}) indicates that $b_{\ell}(q)$ is positive and lower than unity. The behaviour of $c_{\ell}(t;q)$ at late times is deduced from studying the divergence of $\tilde{c}_{\ell}(z;q)$ as $z\to 1^-$. From Eq. (\ref{c_m_tilde}), one has $\tilde{c}_{\ell}(z\to1;q)\simeq \frac{(1-z)^{-(1-b_{\ell}(q))}}{[1-\lambda_{\ell}(1-q)]^{b_{\ell}(q)}}$ which is inverted as
\begin{equation}\label{c_m_asymp}
c_{\ell}(t;q)\simeq \frac{t^{-b_{\ell}(q)}}{\Gamma(1-b_\ell(q))[1-(1-q)\lambda_{\ell}]^{b_{\ell}(q)}} \qquad \text{for}\,\,t\gg1,\; 0<q<1,\; \ell\ge 2.
\end{equation}
This power-law form differs markedly from the exponential decay $\lambda_{\ell}^t$ of the case $q=0$ mentioned previously.
Due to the ordering of the eigenvalues $\{\lambda_{\ell}\}_{\ell=1,\ldots,N}$, it is easy to see from Eq. (\ref{ab}) that, when $0<q<1$, the exponents $b_{\ell}(q)$ are also ordered, with $0<b_2(q)\le b_3(q)\le\ldots\le b_N(q)<1$. Hence, in the sum (\ref{Pij_spect}), the transient part will be controlled by the single mode $\ell=2$ at very large $t$, or
\begin{equation}
	P_{ij}(t)\simeq
	P_j^{\rm (st)}+
	\frac{\left\langle i|\phi_{2}\right\rangle\left\langle\bar{\phi}_{2}|j\right\rangle}{\Gamma(1-b_2(q))[1-(1-q)\lambda_{2}]^{b_{2}(q)}}\, t^{-b_{2}(q)}+\ldots,
 \label{Pij_memory_asymp}
\end{equation}
where we have assumed that the second mode is not degenerate.
In conclusion, $P_{ij}(t)$ tends toward the same stationary distribution $P_j^\infty$ than the Markov process, but the asymptotic approach to this distribution is much slower, in the form of a inverse power-law with exponent $b_{2}(q)$, given by Eq. (\ref{sum_b2}). Furthermore, for finite $t$, in some nodes $j$, the occupation probability in Eq. (\ref{Pij_memory_asymp}) is larger than $P_j^\infty$ while for other nodes, it is lower. In addition, only for $t$ large the result in Eq. (\ref{Pij_memory_asymp}) is a probability satisfying $P_{ij}(t)\geq 0$ and $\sum_{j=1}^N P_{ij}(t)=1$.

\section{Continuous-time formulation}\label{sec:cont}

\subsection{Master equation and solution}

In the continuous-time setting, during a time interval $[t,t+dt]$ the walker standing at node $i$ hops with probability $\gamma dt$ to a n.n. node (with the transition probabilities expressed by the same matrix ${\mathbf W}$ as in the discrete time process) or preferentially resets to a previous node with probability $rdt$. The parameters $\gamma$ and $r$ represent the hopping and resetting rates, respectively. With the complementary probability $1-\gamma dt-rdt$, the walker stays at node $i$. Consequently, the master equation satisfied by the occupation probability of node $j$ at time $t$ (for a walk starting from $i$ at $t=0$) is
\begin{equation}\label{master_cont}
\frac{\partial P_{ij}(t)}{\partial t}=-\gamma P_{ij}(t)+\gamma\sum_{m=1}^N w_{mj}P_{im}(t)-rP_{ij}(t)+\frac{r}{t}\int_0^{t}dt'P_{ij}(t').
\end{equation}
The first two terms of the right-hand-side (rhs) of Eq. (\ref{master_cont}) account for the Markov continuous time random walk dynamics; the third term represents the probability loss due to resetting, while the last term implements the preferential revisit rule, similarly to the non-local term of Eq. (\ref{master1}). Similar equations have been introduced for the study of resetting processes with preferential revisits in continuous time, in the context of free Brownian motion \cite{boyer2017long}, Brownian motion in a potential \cite{boyer2024powerlaw} or active particles \cite{boyer2024active}.
\\[2mm]
We look for solutions of Eq. (\ref{master_cont}) using the same ansatz (\ref{Pij_spect}), i.e., $P_{ij}(t)=\sum_{\ell=1}^N c_{\ell}(t;r)\left\langle i|\phi_{\ell}\right\rangle\left\langle\bar{\phi}_{\ell}|j\right\rangle$. By following a similar route than in the discrete case, we arrive at an integro-differential equation for the coefficient $c_{\ell}(t;r)$,
\begin{equation}\label{eq_c_ell_cont}
\dot{c}_{\ell}(t;r)=-[\gamma(1-\lambda_{\ell})+r]c_{\ell}(t;q)+\frac{r}{t}\int_0^{t}dt'c_{\ell}(t';q),
\end{equation}
which needs to be solved with the same initial condition $c_{\ell}(t=0;r)=1$. Multiplying Eq. (\ref{eq_c_ell_cont}) by $t$ and taking the time derivative, one obtains
\begin{equation}
t\ddot{c}_{\ell}+\{1+t[\gamma(1-\lambda_{\ell})+r]\}\dot{c}_{\ell}+\gamma(1-\lambda_{\ell})c_{\ell}=0.
\end{equation}
By making the change of variable $z=-[\gamma(1-\lambda_{\ell})+r]t$, this equation can be rewritten as a confluent hypergeometric equation \cite{abramowitz1965handbook},
\begin{equation}\label{hyper}
z\frac{d^2f(z)}{dz^2}+(b-z)\frac{df(z)}{dz}-af(z)=0
\end{equation}
with parameters $a=\gamma(1-\lambda_{\ell})/[\gamma(1-\lambda_{\ell})+r]$ and $b=1$. The general solution of Eq. (\ref{hyper}) is $f(z)=k_1M(a,b,z)+k_2U(a,b,z)$ where $M$ and $U$ are the confluent hypergeometric functions \cite{abramowitz1965handbook}
and $k_1$ and $k_2$ two constants fixed by the initial condition. The function $U(z)$ diverges as $\ln z$ as $z\rightarrow0$ (or $t\rightarrow0$) and is therefore not acceptable, implying $k_2=0$. From the initial condition (\ref{ini_c_ell}) one gets $k_1M(a,b,0)=1$, while $M(a,b,0)=1$ from its definition in Eq. (\ref{M_def.1}). Hence $k_1=1$ and the solution is simply $c_{\ell}(t;r)=M(a,b,z)$, or
\begin{equation}\label{c_ell_cont}
c_{\ell}(t;r)=M\left( \frac{\gamma(1-\lambda_{\ell})}{\gamma(1-\lambda_{\ell})+r},\, 1,\, -[\gamma(1-\lambda_{\ell})+r]t\right).
\end{equation}
One first notices that by setting $r=0$ in Eq. (\ref{c_ell_cont}) and using the identity $M(1,1,z)=e^z$, one recovers $c_{\ell}(t;r=0)=e^{-\gamma(1-\lambda_{\ell})t}$, the result for the memory-less random walk on the network.
\\[2mm]
We also note that, since $\lambda_1=1$, equation (\ref{c_ell_cont}) gives $c_{\ell=1}(t;r)=M(0,1,-rt)=1$ [see Eq. (\ref{M_def.1})] independently of the resetting rate $r$. Therefore the mode $\ell=1$ is stationary even in the presence of memory, like in the discrete time model. Inserting Eq. (\ref{c_ell_cont}) into the ansatz (\ref{Pij_spect}), one obtains the exact result for all $t$,
\begin{equation}
P_{ij}(t)= \langle i|\phi_{1}\rangle\langle\bar{\phi}_{1}|j\rangle+
\sum_{\ell=2}^N M\left( \frac{\gamma(1-\lambda_{\ell})}{\gamma(1-\lambda_{\ell})+r},\, 1,\, -[\gamma(1-\lambda_{\ell})+r]t\right) \langle i|\phi_{\ell}\rangle\langle\bar{\phi}_{\ell}|j\rangle ,
\label{P_ij_cont}
\end{equation}
as announced in Eqs. (\ref{sum_sta}) and (\ref{sum_tr_cont}). At late times, the form of the transient part corresponding to the terms $\ell\ge 2$ is obtained from the behaviour of $M(a,b,z)$ in the limit $z\rightarrow-\infty$, which is given by \cite{abramowitz1965handbook}
\begin{equation}
M(a,b,z)\simeq \frac{\Gamma(b)}{\Gamma(b-a)}\,|z|^{-a},
\end{equation}
if $a<b$. We deduce from Eq. (\ref{c_ell_cont}) that the coefficients $c_{\ell}(t;r)$ decay algebraically with time, if $(r+\gamma)t\gg 1$,
\begin{equation}
c_{\ell}(t;r)\simeq \frac{[\gamma(1-\lambda_{\ell})+r]^{-\theta_{\ell}(r)}}{\Gamma(1-\theta_{\ell}(r))}\,t^{-\theta_{\ell}(r)},
\end{equation}
and the exponent $\theta_{\ell}(r)$ is given by
\begin{equation}\label{thetaell}
\theta_{\ell}(r)=\frac{\gamma(1-\lambda_{\ell})}{\gamma(1-\lambda_{\ell})+r}\, .
\end{equation}
Since the eigenvalues $\{\lambda_{\ell}\}_{\ell=1,\ldots,N}$ are ordered from largest to smallest, the exponents $\{\theta_{\ell}\}_{\ell=1,\ldots,N}$ defined by  Eq. (\ref{thetaell}) are ordered from smallest to largest: $\theta_1=0<\theta_2<\ldots$. Consequently, at very large $t$, the relaxation of the occupation probability toward the stationary state will be controlled by the mode $\ell=2$, whose exponent is $\theta_2(r)$. If this eigenmode is not degenerate,
\begin{equation}\label{Pij_cont_latet}
P_{ij}(t)= \langle i|\phi_{1}\rangle\langle\bar{\phi}_{1}|j\rangle+
\frac{\langle i|\phi_{2}\rangle\langle\bar{\phi}_{2}|j\rangle [\gamma(1-\lambda_{2})+r]^{-\theta_{2}(r)}}{\Gamma(1-\theta_{2}(r))}\,t^{-\theta_{\ell}(r)}+\dots,
\end{equation}
with $\theta_2(r)$ given by Eq. (\ref{theta2}). The expression (\ref{Pij_cont_latet}) above is quite similar to Eq. (\ref{Pij_memory_asymp}) for the discrete case.

\subsection{Detailed balance}
We conclude these derivations of the main results by commenting that in both the continuous and discrete time approaches, the process fulfills local detailed balance in the limit $t\rightarrow\infty$. Therefore, the stationary density $P_{j}^{({\rm st})}$ in Eqs. (\ref{sum_sta}) or (\ref{sta2}) is also an {\it equilibrium} distribution, like for the standard memory-less random walk. This differs markedly from processes with resetting to a single point, a situation where detailed balance is violated: in that case the asymptotic distribution is different and corresponds to a non-equilibrium steady state (see, e.g., Refs. \cite{ResetNetworks_PRE2020} for random walks on networks and \cite{evans2011diffusion,evans2011bdiffusion,evans2020stochastic,evans2014diffusion} for diffusive systems on the line or in $\mathbb{R}^d$). The breakdown of detailed balance in resetting processes to a single point follows from the fact that there exist a finite probability to jump during one time-step from any site to the resetting point (by construction), whereas in the same time interval, the reverse jump is in general not possible. Therefore, the resetting sites acts as a source of probability fluxes. As shown below, this is no longer the case in the present memory process, where all the nodes are potentially resetting nodes.
\\[2mm]
Let us consider the probability flux $\Phi_{nm}^{(r)}(t)$ caused at time $t$ by resetting between two arbitrary nodes $n$ and $m$. We have $\Phi_{nm}^{(r)}(t)=P_n(t)w^{(r)}_{nm}(t)$, where $P_n(t)$ is the probability of occupying node $n$ (the initial condition is not specified) and $w^{(r)}_{nm}(t)$ the transition rate from $n$ to $m$ through memory. Let us denote as $t_n(t)$ and $t_m(t)$ the total times spent by the walker on the nodes $n$ and $m$ until time $t$, respectively. At late time, $P_n(t)\simeq t_n(t)/t$ if the process is ergodic, while $w_{nm}^{(r)}(t)$ is exactly given by $r t_m(t)/t$ from the preferential revisit rule. Therefore,   $\Phi_{nm}^{(r)}(t)$  tends to $\Phi_{nm}^{(r)}=\lim_{t\rightarrow\infty} rt_n(t)t_m(t)/t^2$, which by symmetry is also the limit of the reverse flux $\Phi_{mn}^{(r)}(t)$ from $m$ to $n$, implying detailed balance.
\\[2mm]
As shown by the argument above, detailed balance has a unusual origin here: it is a consequence of the linearity of the preferential rule. More general memory kernels where recently considered in \cite{boyer2024diffusion} for a $1d$ particle confined by a potential, a problem with a phenomenology similar to the one of the present study. In \cite{boyer2024diffusion}, the relocation times $t'$ in the past were chosen with an arbitrary distribution (not necessarily uniform unlike in Section \ref{sec:disc}). When $t'$ was mostly localized near the initial time, the situation was similar to a standard resetting or restart process, with a non-trivial non-equilibrium steady state violating detailed balance. When $t'$ was distributed broadly over the whole interval $[0,t]$, or peaked near the present time $t$, the Boltzmann-Gibbs distribution was recovered. In the latter case, however, it was not possible to show whether detailed balance held in general.

\section{Examples} \label{sec:examples}

In this section, we apply the above results to a few practical examples to investigate how the network structure affects the spreading of the memory process. In the cases considered, the random walk spectra $\{\lambda_{\ell}, |\phi_{\ell}\rangle, \langle\bar{\phi}_{\ell}| \}_{\ell=1,\ldots,N}$ can be obtained either analytically or by numerical diagonalisation of the matrix ${\mathbf W}$. In the following, we focus on the discrete time approach.

\subsection{Complete graph}\label{sec:complete}

This network is simple yet non-trivial. The sum in the exact expression (\ref{charPij}) for the characteristic function can be performed explicitly and inverted in the large $t$ regime. When each node is connected to the $N-1$ other nodes (self-connections are prohibited hence the trace of ${\mathbf W}$ is zero), there are only two eigenvalues: $\lambda_1=1$ and $\lambda_{2}=-1/(N-1)$, the latter being degenerate. Eq. (\ref{charPij}) becomes
\begin{equation}
\widetilde{P}_{ij}(z)=
\frac{\langle i|\phi_{1}\rangle \langle\bar{\phi}_{1}|j\rangle}{1-z}
+(1-z)^{-(1-b_{2}(q))}[1-(1-q)\lambda_{2} z]^{-b_{2}(q)}
\sum_{\ell=2}^N 
\left\langle i|\phi_{\ell}\right\rangle\left\langle\bar{\phi}_{\ell}|j\right\rangle.
\end{equation}
Since $|i\rangle$ and $|j\rangle$ are orthogonal, 
\begin{equation}
\sum_{\ell=2}^N 
\left\langle i|\phi_{\ell}\right\rangle\left\langle\bar{\phi}_{\ell}|j\right\rangle=\delta_{ij}-\langle i|\phi_{1}\rangle \langle\bar{\phi}_{1}|j\rangle,
\end{equation}
and, since the asymptotic occupation probability is uniform for a graph where all the nodes have the same degree,
\begin{equation}
\langle i|\phi_{1}\rangle \langle\bar{\phi}_{1}|j\rangle=\frac{1}{N}.
\end{equation}
One obtains from these considerations,
\begin{equation}
\widetilde{P}_{ij}(z)=
\frac{1}{N(1-z)}
+(1-z)^{-(1-b_{2}(q))}[1-(1-q)\lambda_{2} z]^{-b_{2}(q)}\left(\delta_{ij}-\frac{1}{N} \right).
\end{equation}
If the number of nodes is fixed but large ($N\gg 1$), then $|\lambda_2|\ll 1$ and 
\begin{equation}\label{b2compl}
b_2(q)\simeq 1-q , 
\end{equation}
from Eq. (\ref{sum_b2}). Using $[1-(1-q)\lambda_{2} z]^{-b_{2}(q)}\simeq 1$ at small $\lambda_2$ gives
\begin{equation}
\widetilde{P}_{ij}(z)\simeq
\frac{1}{N(1-z)}
+\frac{1}{(1-z)^q}\left(\delta_{ij}-\frac{1}{N} \right), \quad N\gg1,
\end{equation}
yielding the large $t$ behavior
\begin{equation}
P_{ij}(t)\simeq\frac{1}{N}+
\frac{t^{-(1-q)}}{\Gamma(q)}
\left(\delta_{ij}-\frac{1}{N} \right), \quad N\gg 1.
\end{equation}
The asymptotic occupation probability of the starting site is thus given by
\begin{equation}\label{completeii}
P_{ii}(t)\simeq\frac{1}{N}+
\frac{t^{-(1-q)}}{\Gamma(q)}
\left(1-\frac{1}{N} \right)\simeq \frac{1}{N}+
\frac{t^{-(1-q)}}{\Gamma(q)},
\end{equation}
and that of any other node $j\neq i$ by
\begin{equation}\label{completeineqj}
P_{ij}(t)\simeq\frac{1}{N}-
\frac{t^{-(1-q)}}{N\Gamma(q)}.
\end{equation}
As expected, these probabilities slowly converge toward $1/N$, with a transient part given by a single power-law whose exponent depends on the memory parameter $q>0$ in a simple way. The stronger the memory, the slower the convergence.

\subsection{Rings}\label{sec:rings}

Our second example is the one-dimensional lattice of $N$ nodes with n.n. connections and a periodic boundary condition (ring) where the node labeled by $N$ is connected to the nodes $N-1$ and $1$. In this case, the matrix ${\mathbf W}$ has eigenvalues $\lambda_{\ell}=\cos[2\pi(\ell-1)/N]$ and right eigenvectors with components $\langle i|\phi_{\ell}\rangle=N^{-1/2}\exp[-{\rm i}2\pi(\ell-1)(i-1)/N]$, while $\langle \bar{\phi}_{\ell}|i\rangle=\langle i|\phi_{\ell}\rangle^*$ (with ${\rm i}^2=-1$ and ${\rm i}^*=-{\rm i}$) \cite{van2023graph,riascos2015fractional}. We obtain $P_{ij}(t)$ by exact enumeration from the identity (\ref{Pij_spect}), where the coefficients $c_{\ell}(t;q)$ are calculated numerically by using Eq. (\ref{cm_master}) recursively for each $\ell$ starting from the initial condition $c_{\ell}(t=0;q)=1$. The initial node is chosen to be $i=1$ in all the examples below, unless indicated.
\begin{figure}[t]
\centering
\includegraphics[width=1.0\textwidth]{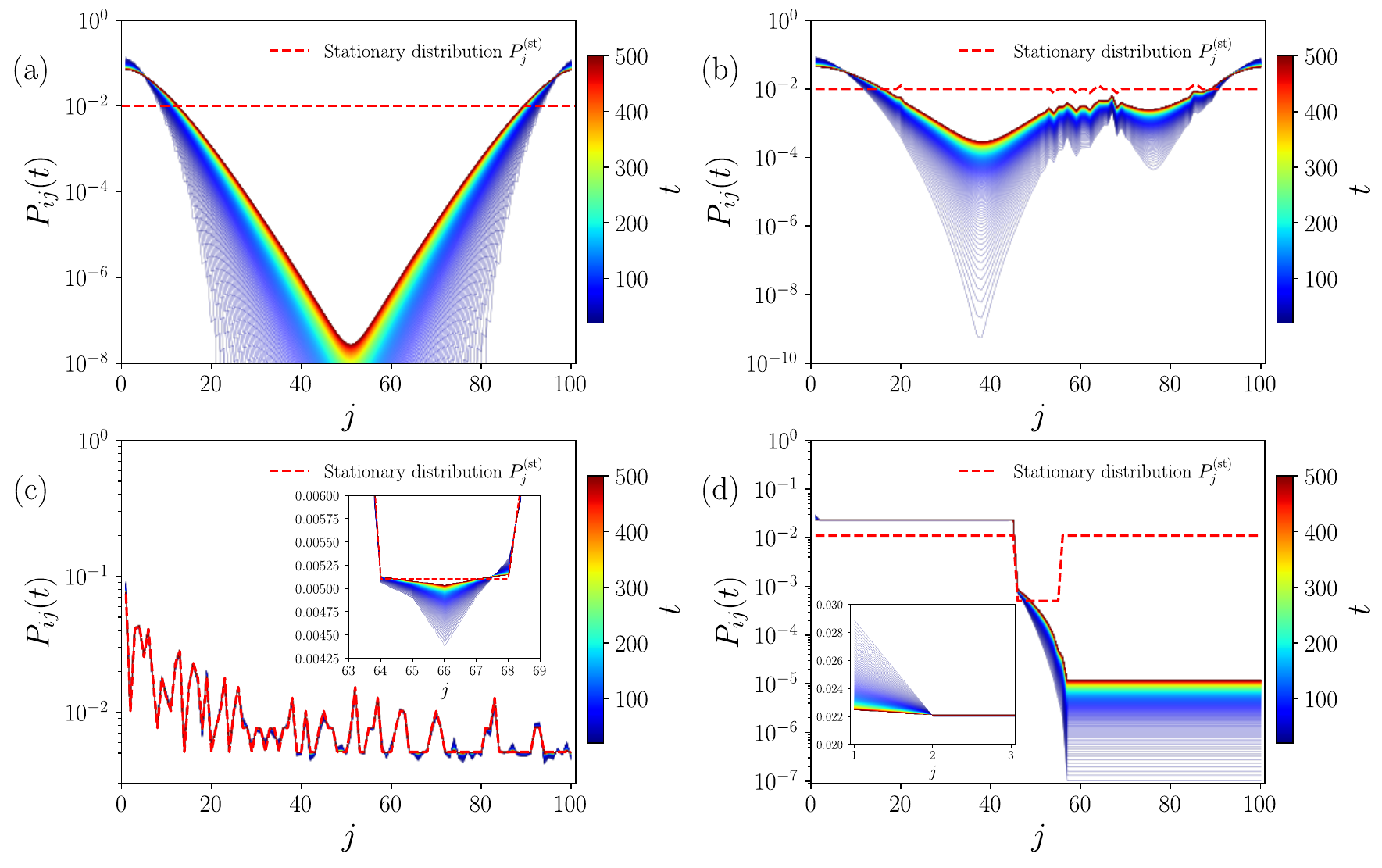}
\caption{Occupation probability at the node $j$, for a walker starting at node $i=1$ in the discrete-time model with memory parameter $q=0.1$, on
different networks of size $N=100$. Time varies as indicated in the color code, with blue and red referring to early and late times, respectively. The dotted lines correspond to the asymptotic stationary distribution in each case. (a) $1d$ lattice with n.n. connections and periodic boundary conditions (ring). (b) Watts-Strogatz network constructed by  randomly rewiring the links of a ring (with n.n. and n.n.n. connections) with probability $p=0.02$. (c) Barab\'asi-Albert network with parameter $m=2$ \cite{barabasi1999emergence}. (d) Barbell graph, composed of two complete networks of 45 nodes, linked to each other by a chain of 10 nodes.}
\label{fig.Pij}
\end{figure}
\\[2mm]
Figure \ref{fig.Pij}a displays the exact $P_{ij}(t)$ obtained as a function of $j$ for a ring of $N=100$ nodes, from early times up to $t=5000$, with the memory parameter set to $q=0.1$. The evolution toward the uniform steady state $P_j^{(\rm st)}=1/N$ becomes extremely sluggish at large times. As shown by Fig. \ref{fig.relaxj}a, the quantity $|P_{ij}^{\rm (tr)}(t)|$ (averaged over all the nodes $j=1,\ldots,N$) barely decays with time.
This can be explained by the fact that
in this example, $\lambda_2=\cos(2\pi/N)$ is very close to unity and $q$ rather large. According to Eq. (\ref{sum_b2}), the exponent of the leading power-law reads
\begin{equation}\label{b2ring}
b_2(q)=\frac{(1-q)[1-\cos(2\pi/N)]}{1-(1-q)\cos(2\pi/N)}\simeq \frac{2(1-q)\pi^2}{qN^2+2(1-q)\pi^2},
\end{equation}
where we have taken $N\gg 1$ in the second equality. With the parameters of Figs. \ref{fig.Pij}a or \ref{fig.relaxj}a, Eq. (\ref{b2ring}) gives the very small exponent value of $b_2(q)=0.0174\ldots$ . Nevertheless, this exponent can be made much larger and, actually, tuned to any value in the range $(0,1)$ by decreasing $q$. For $q\ll1$ (and $N$ still $\gg1$), Eq. (\ref{b2ring}) can be rewritten as 
\begin{equation}\label{b2Brown}
b_2(q)\simeq \frac{2\pi^2}{qN^2+2\pi^2}= \frac{D\pi^2/L^2}{q+D\pi^2/L^2}
\end{equation}
where $D=1/2$ is the diffusion coefficient of the random walk on the lattice with unit spacing, and $L=N/2$. One recovers the expression of the exponent $b_2(q)$ derived by a quite different method in \cite{boyer2024powerlaw} for a Brownian particle with diffusivity $D$ and resetting rate $q$, confined in an interval of length $L$ with reflective boundary conditions. 
\\[2mm]
To check the validity of our predictions, we have performed stochastic simulations of the model in discrete time on the ring and have studied the relaxation of the first moment, a global quantity that is easier to compute than $P_{ij}(t)$ itself (see Fig. \ref{fig.relaxj}b and caption). By setting $q=4\pi^2/N^2$ in the first formula of Eq. (\ref{b2Brown}), one expects a power-law with exponent $b_2(q)\simeq 1/3$: this value actually is in excellent agreement with the simulation results, which exhibit the scaling law over nearly 4 decades. Similarly, the larger resetting probability $q=6\pi^2/N^2$, yields $b_2(q)\simeq 1/4$, also in very good accord with the simulations. These dynamics are clearly much slower than in the reference case $q=0$, whose evolution is exponential (Fig. \ref{fig.relaxj}b).

\begin{figure}[t]
\centering
\includegraphics[width=1.0\textwidth]{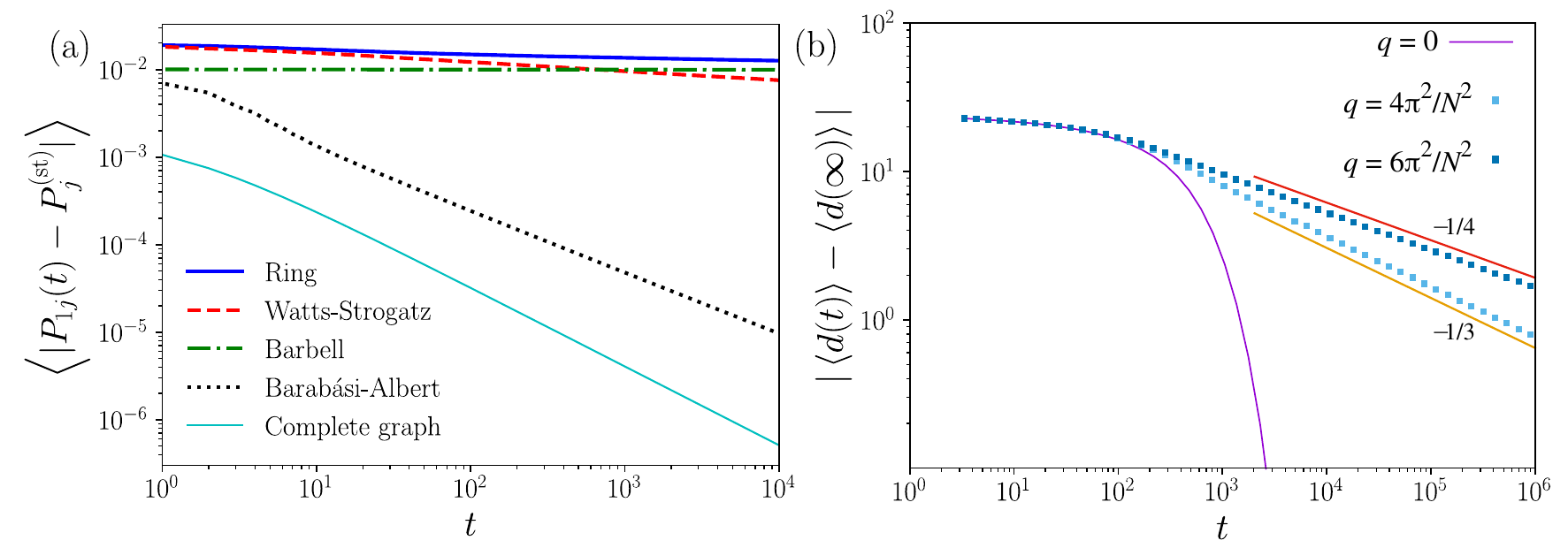}
\caption{Effects of memory in different network topologies. (a) Relaxation of the exact occupation probability toward the stationary distribution, for the complete graph and the different networks considered in Fig. \ref{fig.Pij}, with $q=0.1$ and $N=100$ in all cases (discrete time model). The distance $|P_{ij}(t)-P^{(\rm st)}_j|$ is averaged over the nodes $j$ of each network and the initial node is $i=1$. (b) Relaxation of the first moment of the walker's position toward its asymptotic value on the ring of $N=100$ nodes, obtained from Monte Carlo simulations of the model in discrete time, with $q$ taking the values $6\pi^2/N^2(=0.00592)$, $4\pi^2/N^2(=0.00394)$ and $0$. The first moment is defined by the average $\langle d\rangle(t)=\langle |x(t)-x_{com}|\rangle$, where $1\le x(t)\le N$ is the label of the node occupied by the walker at time $t$ and $x_{com}=(N+1)/2$ the center of mass of the lattice. Averages are performed over $10^5$ independent trajectories and the initial node is $i=N/2$. The asymptotic value of the first moment, $\langle d\rangle(t=\infty)$, is set to its theoretical value $N/4$, obtained by assuming that each node is occupied with equal probability at $t=\infty$. The straight lines are guides to the eye and their slope is given by Eq. (\ref{b2Brown}).}
\label{fig.relaxj}
\end{figure}

\subsection{Watts-Strogatz and Barab\'asi-Albert networks}\label{sec:complex}

We construct a Watts-Strogatz network from a ring of size $N=100$ where the nodes have n.n. and n.n.n. (next-nearest-neighbors) connections, i.e., with degrees $k_i=4$ for all $i$. The links are rewired to randomly chosen nodes with probability $p=0.02$, creating a disordered network with the small-world property \cite{watts1998collective}. Fig. \ref{fig.Pij}b, shows $P_{ij}(t)$ obtained with the same method described in Section \ref{sec:rings}, for a particular realization of the network and where the diagonalization of ${\mathbf W}$ is performed numerically. Due to shortcuts, the relaxation is faster than for the ring, but is still very slow at $q=0.1$. For this network, one obtains $\lambda_2=0.993\ldots$, which yields $b_2(q=0.1)=0.0592\ldots$ from Eq. (\ref{sum_b2}).
\\[2mm]
In contrast to the two previous examples but similarly to the complete graph, Barab\'asi-Albert networks allow a relatively fast diffusion of the random walk with the same memory parameter $q=0.1$. These networks are constructed by the successive addition of nodes that attach preferentially to $m$ previous nodes \cite{boyer2014solvable}. They are scale-free and contain hubs known to favor the spread of random walks with $q=0$. In the example of Fig. \ref{fig.Pij}c, the nodes $j=1,\ldots,N$ are ordered from earliest to latest and the walker starts from the root node $i=1$. The distribution $P_{ij}(t)$ at $t=5000$ is fairly close to the asymptotic limit $P_j^{(\rm st)}$, although some slow relaxation is still taking place, as illustrated in the inset of Fig. \ref{fig.Pij}c. For this network realization, one has $\lambda_2=0.812\ldots$ therefore $b_2(q=0.1)=0.628\ldots$, an exponent smaller than that of the complete network, where $b_2(q=0.1)\simeq 0.9$ from Eq. (\ref{b2compl}). Fig. \ref{fig.relaxj}a clearly shows the power-law decay of $|P_{ij}^{\rm tr}(t)|$ in these two examples, instead of the exponential behavior that characterizes the case $q=0$.

\subsection{Barbell graph}\label{sec:barbell}

A Barbell graph is constructed by connecting two arbitrary connected graphs by a bridge \cite{asmiati2018locating}. Here we choose the two graphs to be complete ones, of $m=45$ nodes each, the bridge being a one-dimensional chain of $n=10$ nodes. This structure mimics two fully connected communities separated intermediate nodes. As shown in Fig. \ref{fig.Pij}d, the relaxation takes place in two phases. The random walker first equilibrates locally and rather rapidly within the community to which the initial node $i=1$ belongs. In the inset of Fig. \ref{fig.Pij}d, the relaxation of $P_{11}(t)$ is visible and occurs roughly as $t^{-(1-q)}$ according to Eq. (\ref{completeii}). Because of the factor $1/m$ in the correcting term of Eq. (\ref{completeineqj}), the other nodes $j\neq 1$ of that community have an occupation probability much closer to $1/(m-1)$. In a second phase, the occupation probability of the other community gradually increases from 0, in a uniform way. For this second phase, the evolution is overall very slow: even at large times, the walker remains practically trapped in the original community ($1-\lambda_2\simeq 0.0001$. These findings suggest that our random walk model could be used to detect communities in networks.
\subsection{ Regular comb}
\begin{figure}[t]
	\centering
	\includegraphics*[width=0.75\textwidth]{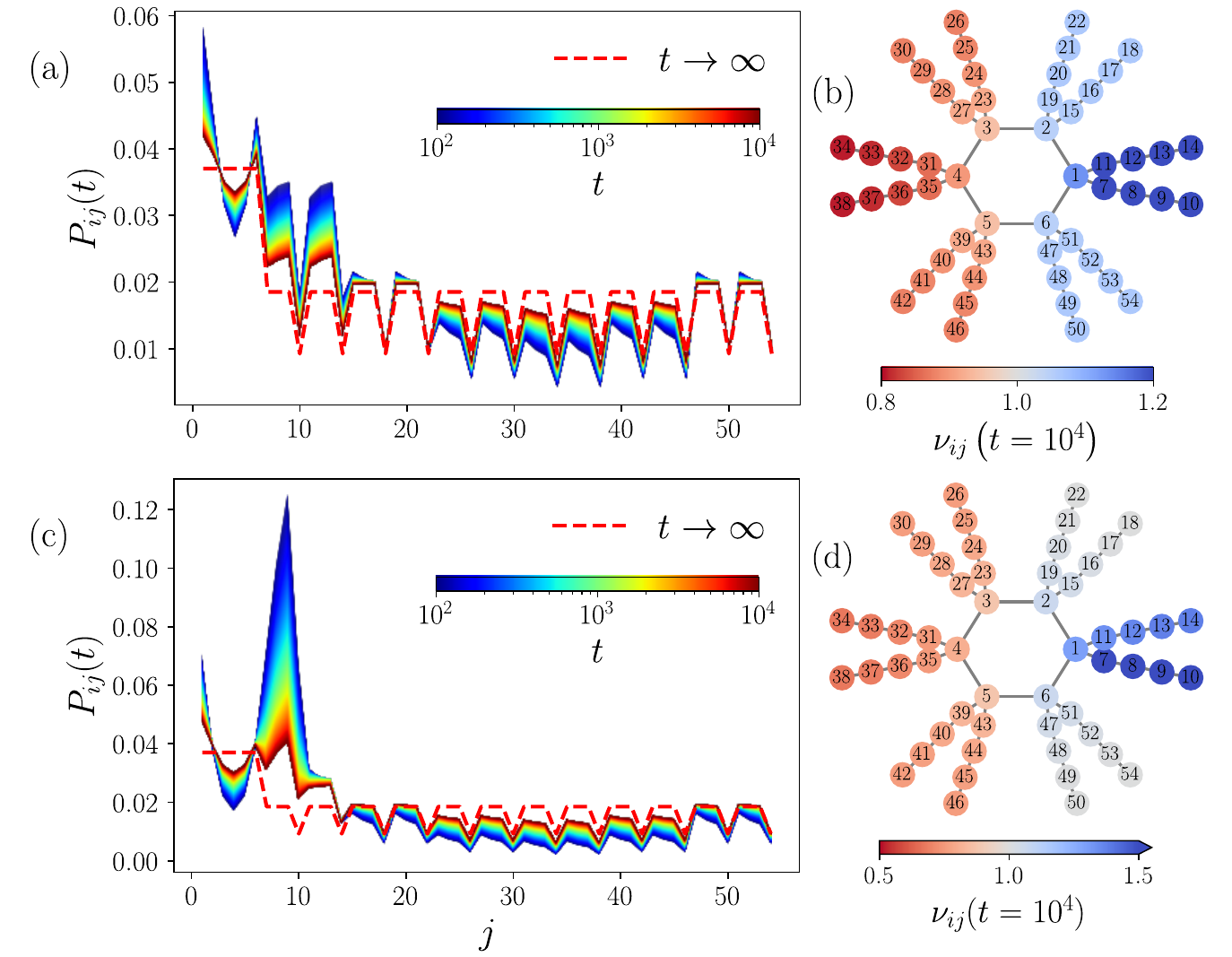}
	\vspace{-2mm}
	\caption{\label{Fig_4} Effects of memory on a regular comb with $L_x=6$ and $L_y=8$. (a) Occupation probability $P_{ij}(t)$ at the node $j$ for a walker starting at $i=1$ (on the ring) and with memory parameter $q=0.1$. The curves depict the results for $t\in [10^2,10^4]$ codified in the color bar, the dashed line shows the stationary distribution $P_j^{(\mathrm{st})}$ obtained in the limit $t\to \infty$. (b) Network analyzed in (a), where the nodes are colored using $\nu_{ij}(t)$ [defined by Eq. (\ref{nu_def})] at $t=10^4$ and encoded in the color bar. Panels (c) and (d) present the same analysis as in (a) and (b), respectively, but for a walker starting at the end of a branch ($i=10$). }
\end{figure}
In this section, we analyze the random walk with memory on a regular comb structure \cite{AgliariPRE2014}. This graph consists of a branched structure derived from a ring of size $L_x$ (assumed even for simplicity), and two side-chains of length $L_y/2$ that are attached to each node of the ring (see Fig. \ref{Fig_4}b and d). The resulting network is a bipartite graph with $N = L_x(L_y + 1)$ nodes. In general, understanding the dynamics of random walkers on comb networks is of significant interest due to their relevance as a simple prototypical model of heterogeneous media with diverse applications in the context of diffusive transport \cite{Durhuus_2006,MENDEZ201346}. Recent studies have focused on topics such as the encounter statistics of simultaneous Markovian random walkers \cite{AgliariPRE2014,PhysRevE.103.042312,Reset_MultipleNodes_2021}, or a single random walk with stochastic resetting \cite{domazetoski2020stochastic,Puigdellosas_2023}, among others.
\\[2mm]
In Fig.  \ref{Fig_4}, we explore the effects of memory for a walker on a regular comb structure with $L_x=6$ and branches of length $L_y/2=4$, the network having $N=54$ nodes in total. In Fig.  \ref{Fig_4}a, we present $P_{ij}(t)$ as a function of the nodes $j$ for a walker with memory parameter $q=0.1$ and starting at a ring node ($i=1$). The results display the occupation probabilities for $t$ spanning the interval $10^2\leq t \leq 10^4$. Our findings show that in the comb structure, the values $P_{ij}(t)$ converge slowly to the stationary distribution $P_j^{(\mathrm{st})}$ obtained in the limit $t\to \infty$. As in the previous examples, this behavior is mainly governed by the eigenvalue $\lambda_2$ of the transition matrix $\mathbf{W}$ which is $\lambda_2=0.9652$ here. In Fig.  \ref{Fig_4}b, we have colored the nodes of the network to quantify how close the occupation probabilities $P_{ij}(t)$ are to their respective asymptotic values $P_j^{(\mathrm{st})}$. To this end, we have represented the quantity
\begin{equation}\label{nu_def}
	\nu_{ij}(t)\equiv\frac{P_{ij}(t)}{P_j^{(\mathrm{st})}},
\end{equation}
and the colors codify $\nu_{ij}(t=10^4)$ for each $j=1,\ldots,54$. The numerical values show how the effects of memory are distributed in the network with values $\nu_{ij}(t=10^4)>1$ for nodes closer to the initial node $i=1$ and $\nu_{ij}(t=10^4)<1$ for the nodes $j=3,4,5$ and the branches connected to these nodes. Interestingly, the nodes of the branches connected to the starting node $i$ are relatively more visited than the node $i$ itself.
\\[2mm]
In Fig.  \ref{Fig_4}c-d, we repeat the same analysis for the occupation probabilities by changing the initial condition to $i=10$, a node located at the extreme of one of the branches connected to node $1$. The values for $\nu_{ij}(t=10^4)$  in Fig.  \ref{Fig_4}d show an asymmetry between the two branches connected to node 1, with the highest values found for $j=7,8,9,10$. For the rest of the nodes $2,3,\ldots,6$ and their branches, a symmetric behavior is observed. 
\section{Conclusion}\label{sec:concl}

We have studied a random walk model with long-range memory on arbitrary finite networks for which the occupation probability can be obtained analytically. The model consists of a standard random walk evolution which is interspersed in time with stochastic resetting events to nodes visited in the past. These revisits by memory are linearly reinforced, i.e., a given previous node is revisited with a probability proportional to the total amount of time spent there by the walker since the initial time. Whereas standard random walks relax exponentially in time toward their stationary equilibrium distribution, relaxation is anomalous in this model. As soon as the resetting probability is non-zero, the approach toward the steady state takes the form of an inverse power-law asymptotically. Nevertheless, the stationary distribution remains the same equilibrium distribution as without memory. Due to the frequent revisits to nodes that are often visited (in the vicinity of the starting node, initially), the exploration of the network by the walker is slow and does not involve a characteristic time-scale: in this sense, it is critically self-organized. Scale invariance is a consequence of the preferential revisit mechanism and of the discrete nature of the spectrum of the underlying random walk matrix.
\\[2mm]
This study extends to networks a previous continuous space version of the model consisting in a $1d$ Brownian particle confined by an arbitrary potential \cite{boyer2024powerlaw}. This problem also exhibited a self-organized critical relaxation controlled by an exponent that depended on the resetting rate and on the smallest non-zero eigenvalue associated to the relaxation of the memory-less particle in the potential. Meanwhile, the stationary state of the particle with memory was still given by the Boltzmann-Gibbs distribution and fulfilled detailed balance despite resetting. Therefore, the phenomenology of processes with preferential revisits in bounded spaces seems to be quite general. In comparison, when the particle moves unbounded on the line \cite{boyer2017long} or on an infinite lattice \cite {boyer2014random}, no stationary state exists and diffusion is logarithmic in time.
\\[2mm] 
Power-law relaxation dynamics toward equilibrium states were also found in other confined non-Markov processes, such as continuous time random walks in an external potential and with diverging waiting times \cite{metzler2000random}. In these diffusion processes of very different nature from the ones considered here, all the eigenstates decayed with the same exponent and thus contributed to the relaxation. This exponent was independent of the potential and was equal to the index of the power-law of the waiting time distribution.
\\[2mm]
A key feature of the preferential relocation models lies in the fact that their evolution can be expressed in terms of the solution of the underlying memory-less process, as exemplified by Eq. (\ref{sum_tr_cont}). Hence the approach is quite flexible and could be applied to other types of random walks on networks, for instance those considering local biases depending on the degree of the nodes, or non-local jumps. The ability of processes with preferential revisits to find target nodes in search tasks should also be investigated in future studies. Unfortunately, a formal calculation of the mean first passage time seems very difficult to achieve and this issue remains outstanding. The framework exposed here could be applied to the study of human mobility, a topic where the use of principles based on preferential revisits to previous locations is customary \cite{song2010modelling,pappalardo2016human,wang2019extended,schlapfer2021universal,cabanas2023human}, and to epidemic spreading \cite{Bestehorn_PRE2022}.
\\[5mm]
{\bf Acknowledgements:}
We thank S. N. Majumdar for many fruitful discussions. DB and APR acknowledges support from Ciencia de Frontera 2019 Grant 10872 (Conahcyt, Mexico).
\\[5mm]
{\bf Data Availability Statement:} The data that support the findings of this study are available from the corresponding author upon reasonable request.
\\[5mm]
{\bf Author Declarations:} The authors have no conflicts to disclose.



\providecommand{\noopsort}[1]{}\providecommand{\singleletter}[1]{#1}%

\end{document}